\documentclass[10pt,a4paper]{article}

\usepackage[margin=0.65in]{geometry}
\usepackage[pdftex]{graphicx}
\usepackage{url}
\usepackage{natbib}
\usepackage{authblk}
  
\title{Time-frequency or time-scale representation fission and fusion rules.}
\author[1]{Coen Jonker}
\author[1,2]{Arryon D. Tijsma}
\author[1,2,*]{Ronald A.J. van Elburg}
\affil[1]{Institute of Artificial Intelligence, Faculty of Mathematics and Natural Sciences, University of Groningen}
\affil[2]{SoundAppraisal BV\\  Groningen, The Netherlands}
\affil[*]{Correspondence: RonaldAJ@vanElburg.eu}

\date{}

\begin{document}
\maketitle

\subsection*{Abstract}\label{abstract}

Time-frequency representations are important for the analysis of time
series. We have developed an online time-series analysis system and
equipped it to reliably handle re-alignment in the time-frequency plane.
The system can deal with issues like invalid regions in time-frequency
representations and discontinuities in data transmissions, making it
suitable for on-line processing in real-world situations. In retrospect
the whole problem can be considered to be a generalization of ideas
present in overlap-and-add filtering, but then for time-frequency
representations and including the calculation of non-causal features.
Here we present our design for time-frequency representation fission and
fusion rules. We present these rules in the context of two typical use
cases, which facilitate understanding of the underlying choices.

Topics: time-series, time-frequency representations, time-scale
representations, computational auditory scene analysis, feature fusion,
data fusion, common representational format, libSoundAnnotator.

\section{Introduction}\label{introduction}

When building a system for online time series analysis one runs into the
problem of proper time-frequency alignment. Although this problem by
necessity appears to all involved in building such systems it is not
given much attention in the literature. However, when presenting our
software, libSoundAnnotator, we found it to be a non-trivial aspect of
the software. And although there is a wealth of literature on processing
time series using time-frequency representations there is little
information available on the implementation of systems producing and
fusing these representations in an online time-series analysis system.
It is our impression that those involved in building these systems value
this issue as a small implementation detail, that is certainly how we
saw it initially. However over time, in contacts with peers and
students, it became clear that proper time-frequency alignment is worth
explaining in some detail. 

The literature available on temporal alignment is predominantly aimed at
aligning signals from different sensors. Due to the nature of that
problem it involves statistical methods for inferring the time-frequency
shift needed to align signals \cite{Mitchell2007}. This is not the
problem we want to address here. The problem we want to address is a
much simpler one. Suppose you are processing a time series online and
the processing requires the creation of intermediate time-frequency
representations which need to be combined to calculate another
time-frequency representations. How should you, in this situation, setup
your metadata to make it possible to re-align the representations,
i.e.~how can you create a common representational format
\cite{Mitchell2007} for this scenario. In addition we make use of
non-causal features. Although it is possible to make our features causal
by manually introducing extra delays, we prefer an approach where the
data indicates how it can be aligned. Thus taken together non-causal
time-frequency representation processing is what sets our processing
requirements apart from what is commonly done in signal processing.

In a quick inventory of open source signal processing software we found
signal processing tools for offline analysis and for real time effects.
In, for example, the field of music information retrieval we found:
\begin{itemize}
\item \url{http://librosa.github.io/librosa/}{LibROSA} \cite{McFee2015},
\item \url{http://essentia.upf.edu/}{Essentia} \cite{Bogdanov2013}.
\end{itemize}
Judging from the algorithms provided the music information retrieval
systems are closest to our computational auditory scene analysis goals.
They are aimed at offline use, although Essentia also offers part of
their functionality in a closed source real-time system. In the area of
computer algebra system we found:
\begin{itemize}
\item \url{https://www.gnu.org/software/octave/}{GNU Octave} \cite{Eaton2015},
\item \url{https://cran.r-project.org/web/packages/signal/index.html}{R-signal} \cite{Signal2014},
\item \url{https://www.scilab.org/scilab/features/scilab/signal_processing}{SciLab} \cite{Scilab2012},
\end{itemize}
all providing signal processing capabilities,
just as the Python packages on which we built or implementation:
\begin{itemize}
\item \url{http://www.numpy.org/}{Numpy} \cite{Walt2011},
\item \url{http://www.scipy.org/}{Scipy} \cite{Scipy2016}.
\end{itemize}
For audio
processing aimed at music synthesis we further found:
\begin{itemize}
\item \url{http://faust.grame.fr/}{Faust Programming Language} \cite{Jouvelot2011,Orlarey2004},
\item \url{https://ccrma.stanford.edu/software/stk/}{The Synthesis ToolKit in
C++} \cite{Scavone2005}. 
\end{itemize}
This list is by no means exhaustive, and in
addition open source development is a very dynamic field making an
overview near obsolete at the moment of publishing. Important for our
purposes here, all of the projects above seem to rely on either
processing whole files at once or using overlap-and-add filter
implementation. We have found no indications for any of these
implementations that they work with time-frequency representations and
non-causal features in an online setting.

Originally our need for online processing arose as part of the Sensor
City Sound project, where we were striving to abstract information close
to the microphone. In this way it should become possible to remove
privacy sensitive aspects before data entered the Sensor City Intranet,
effectively creating a smart sensor. As it was upfront unclear what
could be deployed to the available hardware we also chose for a
networked solution where it was assumed that different processing steps
could take place on different hardware. To make this possible the
necessary information for temporal and frequency alignment should be in
the data travelling from processor to processor, and it wasn't possible
to rely on a central data structure for communication between
processors. Relying on a central data structure for alignment was the
approach chosen in an earlier Matlab\footnote{The Mathworks Inc.,
  Natick, Massachusetts, USA} based implementation of an online CASA
system, this system was never published but for some indicative results
obtained with it see \cite{Krijnders2010},
\cite{Krijnders2010a}.

Before we continue a note on terminology: we speak about chunks instead
of frames in most of this paper. In the first processing stages the
chunks in libSoundAnnotator consist of a single frame and its
accompanying metadata. In our intended further development we foresee
that chunks further in the processing chain will contain information at
higher abstraction levels no longer directly related to digital signal
processing. Therefore, we will stick to the word chunk.

\section{Use Cases libSoundAnnotator}\label{use-cases-libsoundannotator}

At present we distinguish two different scenarios from the end user
point of view:

\begin{itemize}
\item
  offline file processing
\item
  online microphone processing
\end{itemize}

We start with the offline file processing use case. This use case can be
used to illustrate time-frequency alignment. The second use case adds
the complications associated with temporary failures somewhere in the
processing chain. In a networked solution this problem can arise through
failure of connectivity, processing and/or acquisition. The system is
designed to maintain correct alignment while such incidents happen.

We used two alignment design principles:

\begin{itemize}
\item
  As much as possible of the merging should be handled by the framework,
  leaving only processor specific data handling to the processor.
\item
  Only valid time steps are published, to keep alignment simple the
  invalid scales are included in the representation, and they can be
  overwritten by the receiving processor to manage problems with zero's
  and NaN values in later calculations.
\end{itemize}

In addition we use the following simplifying assumptions:

\begin{itemize}
\item
  We assume failures are rare. Which allows us to keep the buffering
  mechanism simple. We will simply accept a failure, fast forward to a
  failure free time point and propagate the failure without attempts to
  recover lost information.
\item
  We assume that the chunks contain a sufficiently long time interval
  that there are always valid time values present in each single chunk.
\end{itemize}

For now this last assumption puts a limit on the total length of the
processing chain, or alternatively it puts a lower bound on the length
of the time interval included in a chunk.

\subsection{Use Case I: Estimating the relative amount of tonal energy
represented in a
wav-file.}\label{use-case-i-estimating-the-relative-amount-of-tonal-energy-represented-in-a-wav-file.}

An earlier version of this software was used by \cite{vEA2} to
estimate the distribution of energy over different sound types: pulsal,
tonal, noise. They defined tract features \(T_{-}\) and \(T_{|}\) which
provide an indication of sound structure. In addition they showed how
these correlate with human perception. Here we will review how using
libSoundAnnotator the tonal energy can be calculated per frequency band
and time interval. This function is implemented in the PTN\_Processor,
which extracts energy in pulses (P), tones (T) and noises (N) and in
addition it provides the total energy in the resulting time-frequency
regions. To achieve this the PTN\_Processor receives an energy
representation (E) from a GammaChirp FilterBank (implemented as a set of
FFT based overlap-and-add filters) and it receives tract-features
(\(T_{-}\), \(T_{|}\)) from two StructureExtractor instances. In the
PTN\_Processor the local energy attributable to for example tones is
calculated by thresholding the corresponding tract feature and
multiplying it with the local energy, i.e. the local tonal energy is
given by \(E_{T}(t,f)=E(t,f)\sigma(T_{-}(t,f))\). For the full
computation the following steps are needed: reading from file,
down-sampling, conversion to cochleogram \(E(t,f)\), extraction of
horizontal tract-feature \(T_{-}\) from the cochleogram, multiplication
of energy in cochleogram with sigmoid function of the horizontal tract
feature and areal averaging, saving of the result to file.

File processing is typically done on a single machine while the
processors are distributed over its cores. Several processors are
started each in their own process, and when running on a single machine
they are linked by pipes. Due to the high reliability of these pipes we
can trust that data published by one process will be received by all
subscribed processes, and we don't have to deal with data loss. As long
as we have a linear topology of processing steps arrival of all data
will be in the same order on all processes. However if we are forking
data flow to two processors and subsequently merge data coming from
these two data flow paths then there is no guarantee that data belonging
together will arrive together. In fact this is unlikely if one path
contains an extra processing step while the other path is actually a
direct connection to the merging process. This is actually the case for
our scenario: \(E(t,f)\) is used for calculation of the tract features
and it also merged with these tract-features in the PTN-Processor. In
that case some buffering is needed on the receiving end until data for
the same time-frequency points has arrived through both paths.

\begin{figure}[htbp]
\centering
\includegraphics[width=.5\textwidth]{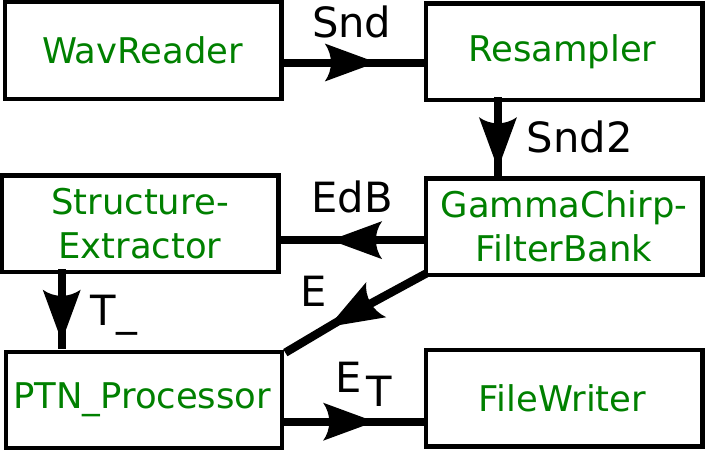}
\caption{Data flow between processors for calculation of tonal energy
\(E_{T}\). Notice that representations needed for the calculation of
\(E_{T}\) reach the PTN\_Processor via two paths thus making a merge
necessary.\label{fig:dataflow}}
\end{figure}

\begin{figure}[htbp]
\centering
\includegraphics[width=\textwidth]{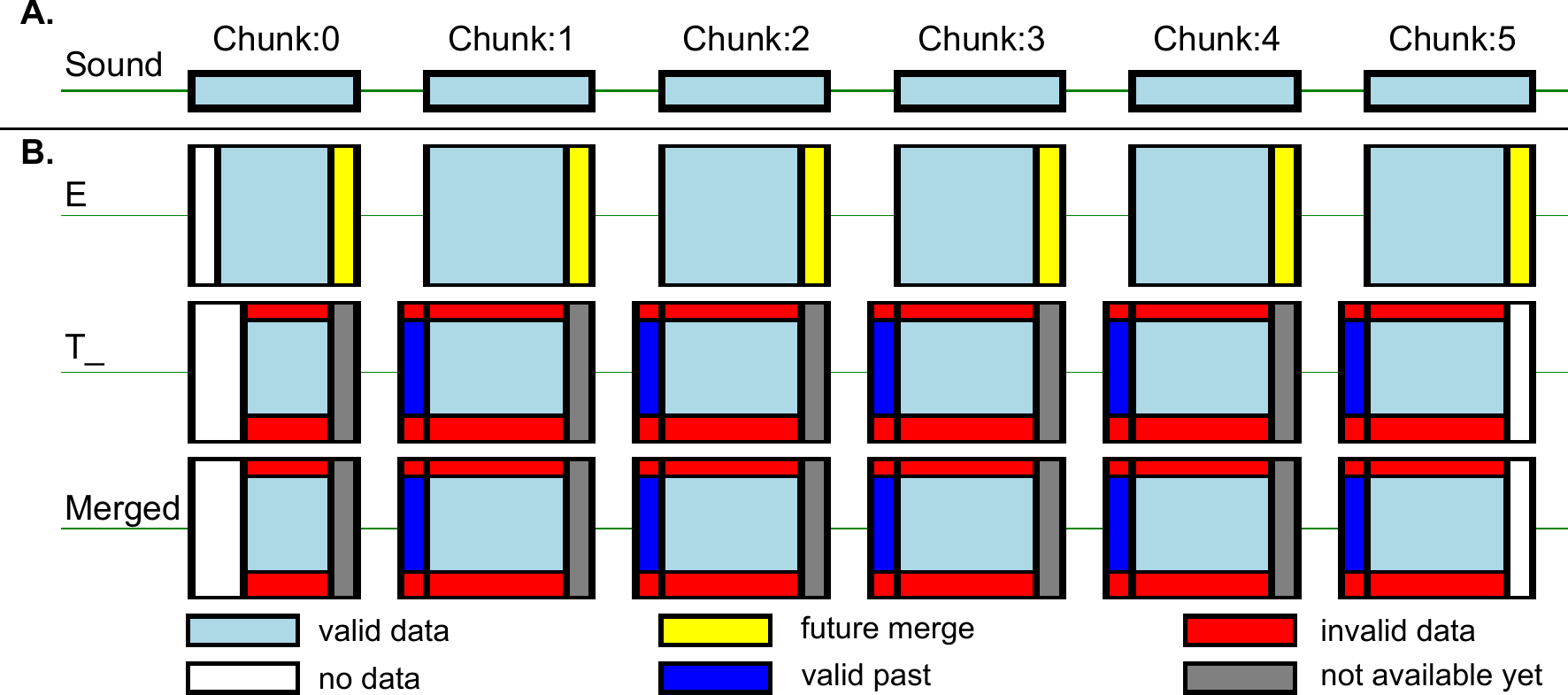}
\caption{Offline file processing: A. Chunks as produced by WavReader, B.
Perspective on chunks at the PTN\_Processor which receives \(E\) and
\(T_{-}\) and calculates
\(E_{T}(t,f)=E(t,f)\sigma(T_{-}(t,f))\).\label{fig:alignfile}}
\end{figure}

To be able to calculate the tonal energy \(E_{T}(t,f)\) for a single
time-frequency location \((t,f)\) requires the presence at that location
of the energy \(E(t,f)\) and the horizontal tract feature
\(T_{-}(t,f)\), see figure \ref{fig:dataflow}. Now the energy at a given
frequency can only be calculated when there is a history available of
the full length of the applied \(\gamma\)-chirp filter, which in our
setting is typically set to 100 ms. Furthermore because the tract
features look at a region surrounding a time-frequency location they are
only defined if that region is completely within the current
E-representation. Thus tract-features are only defined if the
time-frequency location is sufficiently far away from the lowest and
highest energy in the E-representation and is sufficiently far away from
the time for which the first and last received energy are valid. This is
illustrated in figure \ref{fig:alignfile}.A which shows the chunks as
published by the WavReader. The panel \ref{fig:alignfile}.B shows the
PTN-Processor perspective on its incoming \(E\), \(T_{-}\) and outgoing
representations \(E \sigma(T_{-})\). The colors indicate whether data is
present and if present whether it is valid data. We chose to transmit
some invalid data to keep the implementation relatively simple.

The alignment principles in the architecture make sure the data is
aligned before it is handed to a processor. The processor can then
impose its own rules to adjust alignment further. This is to say that a
processor receives only data belonging to the same time-interval. This
data is accompanied by meta-data indicating how it is shifted with
respect to the original time-series. We determined that we need 4
parameters to describe these shifts and invalid regions, where some of
the names also prelude on the next use case:

\begin{itemize}
 
\item
  includedPast (p): This parameter is used for non-causal features or
  their non-causal part. It measures how many time steps from the future
  are used to calculate the feature value at this point. To simplify the
  treatment the maximum value over all frequencies is used.
\item
  droppedAfterDiscontinuity (d): This parameter is used for causal
  features or their causal part. It measures how many time steps from
  the past are used to calculate the feature value at this point. To
  simplify the treatment the maximum value over all frequencies is used.
\item
  invalidLargeScales (l): This parameter is used for features depending
  not only on the frequency considered but also on lower frequencies
  (larger scales). It measures how many higher frequency channels are
  used to calculate the feature value at the point under consideration.
  To simplify the treatment the maximal value is used subject to the
  constraint that only values which lead to frequencies outside the
  TF-representation are considered.
\item
  invalidSmallScales (s): This parameter is used for features depending
  not only on the frequency considered but also on higher frequencies
  (smaller scales). It measures how many lower frequency channels are
  used to calculate the feature value at the point under consideration.
  To simplify the treatment the maximal value is used subject to the
  constraint that only values which lead to frequencies outside the
  TF-representation are considered.
\end{itemize}

All these 4 parameters appear in two forms: a cumulative form in the
meta data of an incoming (merged) chunk and a relative form related to a
processor indicating how the processor will impact alignment of the
features it will publish. Because a single processor can produce several
features it contains alignment parameters for all features produced.
When publishing an outgoing chunk (\texttt{Out}) will get the following
alignment parameters from the incoming (\texttt{Merged}) chunks and the
feature publishing processor (\texttt{Feature}):
\begin{eqnarray*}
Out.p&=&Merged.p+Feature.p\\
Out.d&=&Merged.d+Feature.d\\
Out.l&=&Merged.l+Feature.l\\
Out.s&=&Merged.s+Feature.s
\end{eqnarray*}
The \(Merged\) chunk is obtained from the incoming chunks. For example
if two incoming chunks need to be aligned we need to use different rules
for merging the alignment parameters. A merged chunk (\texttt{Merged})
derived from two incoming chunk (\texttt{In1} and \texttt{In2}) will
have the following alignment parameters:
\begin{eqnarray*}
Merged.p&=&max(In1.p,In2.p)\\
Merged.d&=&max(In1.d,In2.d)\\
Merged.l&=&max(In1.l,In2.l)\\
Merged.s&=&max(In1.s,In2.s)
\end{eqnarray*}
This alignment merging rule basically states that only areas in which
valid data from all incoming chunks overlap will be valid in the merged
chunk. How these rules work out is shown in figure \ref{fig:alignfile}.

Let us discuss figure \ref{fig:alignfile} and the underlying choices in
more detail. The InputProcessor in this case the WavReader produces
chunks of sound, which typically covers several tenths of a second. Each
chunk receives a number indicating the order in which it was entered
into the system. The light blue color for each chunk in
\ref{fig:alignfile}.A indicates that we consider all samples in a chunk
as valid sound data. After two processing steps we arrive at an energy
or cochleogam representation \(E(t,f)\). Because both processing steps,
i.e.~down-sampling and \(\gamma\)-chirp filterbank require a filter of
finite length it is only after processing the combined length of these
filters in sound data that we arrive at valid data. This is indicated
with white areas at the beginning of all affected chunks. The \(E\)
representation also contains yellow areas, these yellow areas are not
intrinsic to the received data but they are related to the merging
process in the PTN\_Processor. Yellow indicates that the receiving
processor needs to store these for use with other data that still has to
arrive. In this case because the tract feature couldn't be calculated
for these values until the next chunk became available. These future
dependent values are indicated by the gray areas in the next line with
chunks. The blue areas are their counter part, they contain data that
has been prepended containing the features that couldn't be calculated
earlier. The red areas are similar to the white areas in that no valid
representation can be calculated for these areas. However the
representation contains invalid values, typically NaN or 0, at the
time-frequency points that fall into these areas. Also contrary to the
white areas the red areas are not dropped from the chunk. Instead they
are kept to facilitate array operations in later processing steps. Their
invalidity can be read off from the alignment parameters
(invalidLargeScales,invalidSmallScales) introduced above.

The last line of chunks represents the output of the merging operation.
The merging operation takes place in the compositeManager. As soon as
all relevant chunks arrived it will pass all TF-representations, in this
case \(E\) and \(T_{-}\), to a processors processData method. The
merging operation above ensures that all TF-representations arriving at
a processor belong to the same time interval and that the frequencies
stay aligned.

\subsection{Use Case II: Estimating the relative amount of tonal energy
present in sound arriving at a
microphone}\label{use-case-ii-estimating-the-relative-amount-of-tonal-energy-present-in-sound-arriving-at-a-microphone}

From the signal processing point of view this is more or less the same
task as in the previous use case. So lets us stress the differences
provided by the context. We analyse here the worst case scenario that
each processor runs on a separate machine, therefore all processors are
connected by network connections and we can expect transmission failures
which affect the processor input. In one context, for example, we used
an indoors wifi connection to relay sound from a microphone to
subsequent processing on a PC and we observed many failures in the
connection. Despite these failures processing on the PC could continue
once the connection was restored without restarting the whole system.
This is possible due to two kinds of metadata, the first kind we already
encountered in use case I the alignment parameters, the second kind we
encounter in this use case is the continuity flag.

The potential continuity flag values are defined in the class
ContinuityMeta as follows:
\begin{verbatim}
# Continuity class: please maintain numerical order, the code assumes
# the mapping is one-to-one.
values = {
    #invalid chunk
    'invalid' : -1,
    #discontinuous subtypes
    'discontinuous' : 0,
    'newfile': 1,
    'calibrationChunk': 2,
    #withprevious subtypes
    'withprevious' : 10,
    'last' : 11
}
\end{verbatim}
Let discuss values in order of appearance. The \texttt{invalid} value is
used to signal that we don't know whether the samples in a chunk are
continuous with the samples in its predecessor or successor or neither.
An example we can give is that some times the microphone input buffer
experiences an overflow, in that case we know there are samples lost and
we have lost continuity between our samples. The \texttt{discontinuous}
value and the other discontinuous subtypes \texttt{newfile} and
\texttt{calibrationchunk} signal that they are discontinuous with their
predecessor. The first chunk from the MicInputProcessor after an
incidental buffer overflow will be flagged as \texttt{discontinuous}.
Subsequent chunks provided by the MicInputProcessor will be marked as
\texttt{withprevious} as they contain frames which are continuous with
those encountered in the previous chunk. There is a special
\texttt{last} value as well which is the counterpart of
\texttt{newfile}, both are used in file processing where it is necessary
to mark the beginning and end of a file. The value
\texttt{calibrationchunk} is a special case and is at the same time a
`discontinuous' and a `last' subtype. However the architecture treats it
as `discontinuous' and leaves it to those processors needing calibration
to take appropriate action on receiving it.

\begin{figure}[htbp]
\centering
\includegraphics[width=\textwidth]{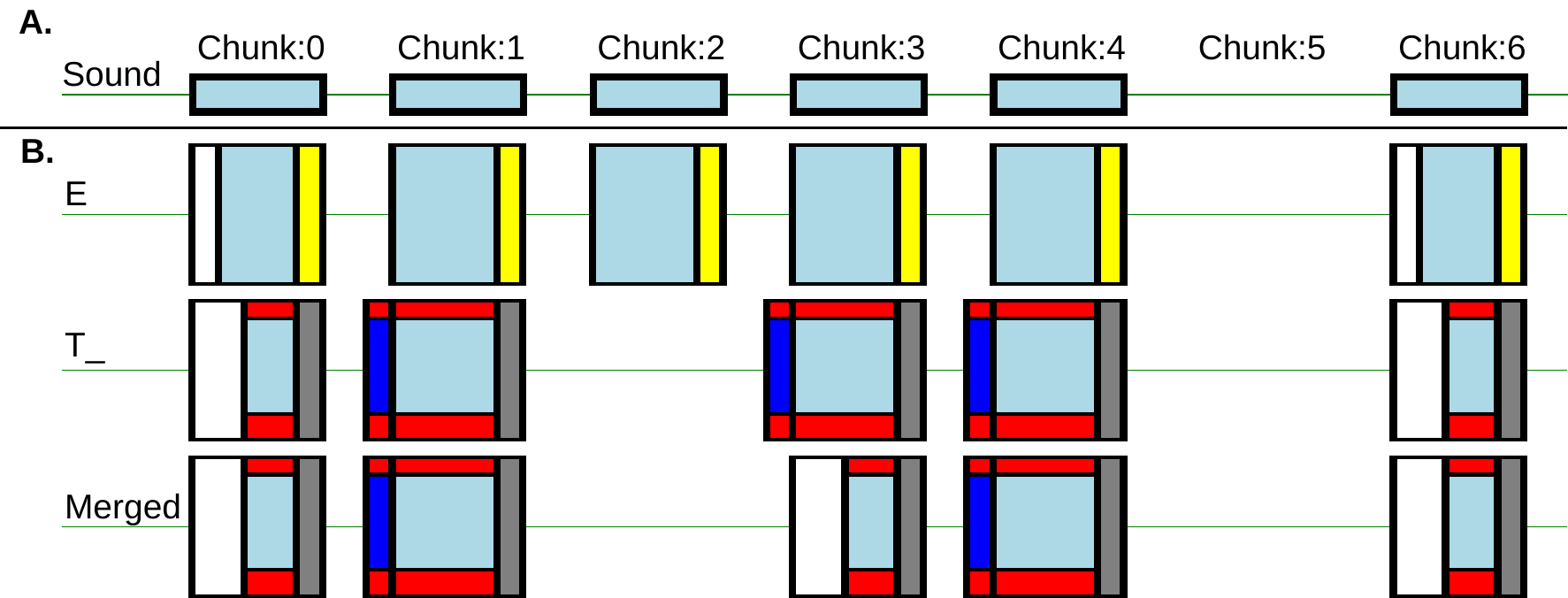}
\caption{Online microphone processing: Impact of discontinuities on
subsequent processing. After a regular start of the system at chunk 0 we
see a transmission failure between a StructureProcessor instance and
PTN\_Processor instance at chunk 2 which results in an absent \(T_{-}\).
And we see how the system recovers with a merged chunk 3. Later on we
see that chunk 5 failed to reach the GammaChirpFilterbank, and we see
how the system recovers with a merged chunk 6.\label{fig:align_mic}}
\end{figure}

Sofar we discussed mainly how the continuity flag on a chunk is set when
obtaining sound from a microphone or file. However, that is not the only
point at which continuity can change, in addition we need to determine
how a continuity change propagates and how it influences subsequent
processing. We have illustrated two of these scenarios in figure
\ref{fig:align_mic}.

Lets discuss the two discontinuity scenarios depicted in figure
\ref{fig:align_mic}. This figure shows two continuity failures. First a
failure in the processing of chunk 2. Chunk 2 is lost during
transmission from the structureExctractor instance to the PTN\_Processor
instance. Second a failure in the data acquisition at the microphone
leads to a buffer overflow during acquisition of chunk 5 which,
therefore, gets flagged as \texttt{invalid}. This invalid chunk is never
published and stays at the MicInputProcessor instance recording the
state of this processor. The next reading of the microphone succeeds and
the MicInputProcessor instance publishes this chunk flagged as
\texttt{discontinuous}.

Lets examine the first scenario a bit more closely. How does it become
clear that \(T_{-}\) was lost. To understand this we need to point out
that we made a simplifying assumption, we assumed that as a rule the
chunks arrive in the correct order. Therefore if chunk \(T_{-,6}\)
arrives we infer that \(T_{-,5}\) was lost. If later on \(T_{-,5}\)
arrives it will be discarded, and we will not rewind history to see if
it can still be fixed. In fact all available chunks with chunk number 5
or lower are discarded after it was established that a discontinuity
occured from chunk 5 to chunk 6. Now on arrival of chunks \(T_{-,6}\)
and \(E_6\), we need to decide which part of the data we will include in
the merge. To keep it simple we discard the information in the buffer,
and try to align data from the currently incoming chunks only. In this
scenario \(T_{-,6}\) and \(E_6\) both have continuity
\texttt{withprevious} despite the discontinuity in the transmission.
Therefore we know that information is available from the start of the
time interval originally inserted as chunk \(Snd_6\) into the processing
chain. Now \(T_{-,6}\) has prepended part of the data which should be
aligned with data coming from \(Snd_5\), this part should be discarded.
Now the length of this part is encoded in the alignment parameter
\texttt{includedPast} \(T_{-,6}.p\) of \(T_{-,6}\). This would be
sufficient if processing stopped here, we will however flag this chunk
as \texttt{discontinuous} and therefore the receiving processor will
assume it does not contain the first part, therefore we use the
parameter \texttt{droppedAfterDiscontinuity} \(Merged.d\) of the \(E\)
merged with \(T_{-}\) to discard even more timesteps. Remember we
assumed these events to be relatively rare, therefore we are convinced
that the benefits of trying to keep things simple outweigh the benefits
of saving additional data. For our wifi failure example this still works
out reasonably well, as the interruptions were often several hours
containing tens of thousands of chunks and then connectivity would
mostly be present for several hours. And thus dropping this data leads
to an increase of data loss less than a promille.

Because of this last choice the number of scenarios to consider when
aligning a chunk from a complete set of received chunks is limited to
three. One scenario for regular continuous operation, which applies if
the resulting merged chunk is of \texttt{withprevious} subtype:

\begin{enumerate}
\def\labelenumi{\arabic{enumi}.}
 
\item
  \textbf{Regular Continuous Operation}: All chunks in the set have
  continuity \texttt{withprevious} or \texttt{last} and the last
  previously completed set of chunks belonged to the directly preceding
  time interval not an earlier one.
\end{enumerate}
And two other cases, which are invoked if conditions are such that after
merging the merged chunk should be of a \texttt{discontinuous} subtype.
This occurs if the last previously completed set of chunks belonged to a
preceding time interval not directly preceding the current interval, or
during a regular startup in which all chunks should be of a
\texttt{discontinuous} subtype.
\begin{enumerate}
\def\labelenumi{\arabic{enumi}.}
\setcounter{enumi}{1}
\item
  \textbf{Regular Discontinuous Operation}: The chunk to be merged is of
  a \texttt{discontinuous} subtype and the resulting merged chunk of a
  \texttt{discontinuous} subtype.
\item
  \textbf{Irregular Discontinuous Operation}: The chunk to be merged is
  of a \texttt{withprevious} subtype and the resulting merged chunk of a
  \texttt{discontinuous} subtype.
\end{enumerate}
All these three scenarios require their own handling of the incoming
data, based on the information contained in the alignment parameters.
Lets link this back to figure \ref{fig:align_mic}. To obtain the merged
data we observe Regular Discontinuous Operation for chunk \(Merged_0\)
at startup, we observe Regular Continuous Operation for chunk
\(Merged_1\) and \(Merged_4\) and we observe Irregular Discontinuous
Operation for the construction of chunks \(Merged_3\). Perhaps somewhat
suprisingly the construction of \(Merged_6\) is obtained through Regular
Discontinuous Operation, this is because the discontinuity arose earlier
in the processing chain. In this case Irregular Discontinuous Operation
occurs when merging takes place for the Resampler, there at the
Resampler discontinuity is spotted through the absence of chunk
\(Snd_5\). Due to our choice to let all discontinuous merged data cover
the same time interval with respect to the original sound data the
Resampler will publish a `regularized' chunk \(Snd2_6\) of the
discontinuous subtype and then after the Resampler all processing for
chunk 6 is regular. This is because the other processors don't receive
\(Snd_6\) directly but instead they receive the `regularized' chunk
\(Snd2_6\) or its regular descendants. Therefore this case illustrates
how `chunk regularization' helps to keep processing simple.

Lets us now describe how in these three scenarios the merged data arrays
are constructed from the incoming data arrays. We will introduce new
notation to illustrate this. We denote the data array in an incoming
chunk by \(A_N^f\), where \(N\) indicates the chunk number, and \(f\)
denotes the continuity subtype and will be replaced by \(d\) for chunks
of the discontinuous subtypes and by \(c\) for chunks of the continuous
with previous chunk subtypes, invalid chunks do not appear in the merge
process as they remain unpublished. The array resulting from the merge
is denoted by \(\tilde{A}_N^f\). Now for the actual merge at array level
we only need to consider a single representation, but to determine the
continuity subtype denoted by \(f\) we need to consider all incoming
chunks and whether the last merged chunk was the direct predecessor of
the current chunk. To achieve this the last merged chunk and the data on
which it was based are kept. And during the merging of all incoming
chunks for chunk number \(N\) we determine \(f\) using the following
rules in pythonesque pseudocode:

\begin{verbatim}
if  not N == N_lastcompleted +1:
    f_merged=d
else:
    if any incoming chunk has f==d:
        f_merged=d
    else:
        f_merged=c
\end{verbatim}

Now that we established the continuity subtype of the merged chunk, we
can select the correct rule to calculate the merged array for each
incoming representation from the rules below:
\begin{enumerate}
\def\labelenumi{}
\item
  \textbf{Regular Continuous Operation}:
  \(\tilde{A}_N^c = A_{N-1}^c(:,e-d_H:) + A_N^c(:,:e-d_H)\)
\item
  \textbf{Regular Discontinuous Operation}:
  \(\tilde{A}_N^d = A_N^d(:,d_L:e-d_H)\)
\item
  \textbf{Irregular Discontinuous Operation}:
  \(\tilde{A}_N^d = A_N^c(:,d_l:e-d_H)\)
\end{enumerate}
Where \(+\) denotes concatenation, between brackets the selection of
part of the arrays is indicated, where we follow the notation used in
Numpy \cite{Walt2011}. The letter \(e\), the first letter from the
word `end', indicates the size of the array for the dimension where it
appears. The parameters \(d_L\), \(d_H\) and \(d_l\) are derived from
the alignment parameters of the incoming chunks, and the alignment
parameters for the merged chunk which have been obtained through
repeated application of the rules outlined earlier for UseCase I. The
\(d\) is derived from the word drop, \(L\), \(l\) and \(H\) refer to low
and high indices respectively:
\begin{eqnarray*}
d_H= \tilde{A}_N.p-A_N.p\\
d_L = \tilde{A}_N.d-A_N.d\\
d_l=\tilde{A}_N.d+A_N.p
\end{eqnarray*}
Where we used the arrays to refer to the corresponding chunk and its
alignment parameters. The rules for time series are obtained by simply
dropping the first dimension representing the frequencies.

\section{Discussion}\label{discussion}

The algorithm presented above provides sound and more general
time-series processing which maintains time-frequency alignment. It can
achieve this even in the presence of incidental acquisition, hardware
and data transmission failures and it that sense the algorithm is
robust.

Our algorithm transmits, at our current standard settings, about 10\%
invalid data, this has been an optimization on developer time. Keeping
this data facilitates the use of standard algorithms from the numerical
python library to do element wise multiplication, and therefore also
reduces the chance of implementation errors. Therefore although the
saving in our standard setting would be significant and in the order of
10\% in computing time, we considered the potential savings not worth
the costs of spending a large amount of developer time at this stage of
development. Similar considerations on developer time and
maintainability in an academic context lead us to transmit
representations using pipes when processes are running on a single
machine, this implies a large amount of data copying. We don't know at
present how this impacts performance, but we imagine that a shared
memory model would eventually be more efficient. What is presented in
this paper does not rely on the exact form in which data is transferred
and can therefore without change be applied to a shared memory model.

There is a part in our design which is design by contract. We made no
provision for working with data coming from two independent sources. In
which case different continuity subtypes can occur on chunks arriving
from the different original sources. That this problem can occur is a
likely signal that the design of continuity can be further improved by
splitting it in two separate flags. One purely aimed at administrating
continuity for use in the distinction between Regular Continuous
Operation, Regular Discontinuous Operation and Irregular Discontinuous
Operation. And another flag indicating the intended continuity from the
perspective of the original source. This being said, scenarios using
different sources need true fusion algorithms, taking into account
deviations in clock speeds, sampling rates and start times, and in our
case a forced alignment of chunk numbering. Our code does not provide
such functionality at present.

The whole library implementation is made available as open source through 
GitHub: \url{https://github.com/soundappraisal/libsoundannotator} and licensed 
under the Apache License, Version 2.0. Both
UseCase are available as documented code examples  also available through GitHub 
\url{https://github.com/soundappraisal/soundannotatordemo} and licensed under 
the Apache License, Version 2.0. 

\section{Funding}\label{funding}

Part of the research described in this paper was conducted as part of
the SensorCity Sound project which was funded by the European Union, the
European Regional Development Fund, the Dutch Ministry of Economic
Affairs, and The Northern Netherlands Provinces Alliance, KOERS NOORD.

\section{Author Contributions}\label{author-contributions}

CJ, RAJvE: Design and implementation streamboard.
AT: Design and implementation networking.
RAJvE: Design and implementation alignment and continuity handling mechanism and overall design, wrote the paper.
AT, CJ, RAJvE: Contributed code.

\bibliographystyle{apalike}

\newpage
\appendix
\section{Appendix: Calibration}\label{appendix-calibration}

The calibration signal is needed in those use cases where a processor
needs to estimate its own internal parameters while taking into account
the input it can expect. Tract-features \cite{vEA2} for example
require knowledge about the correlations distances found when processing
white noise. These correlation distances are dependent on the method and
parameters used to create a time-frequency representation. Calibration
is a necessary step preparatory step in all use cases, from the
implementation point of view it is, however, a special case of the
offline file processing scenario in which the whole file is contained in
a single frame or chunk, in addition this single chunk is flagged as
being a calibration chunk.

\section{Appendix: Streamboard}\label{appendix-streamboard}

\begin{figure}[htbp]
\centering
\includegraphics[width=.7\textwidth]{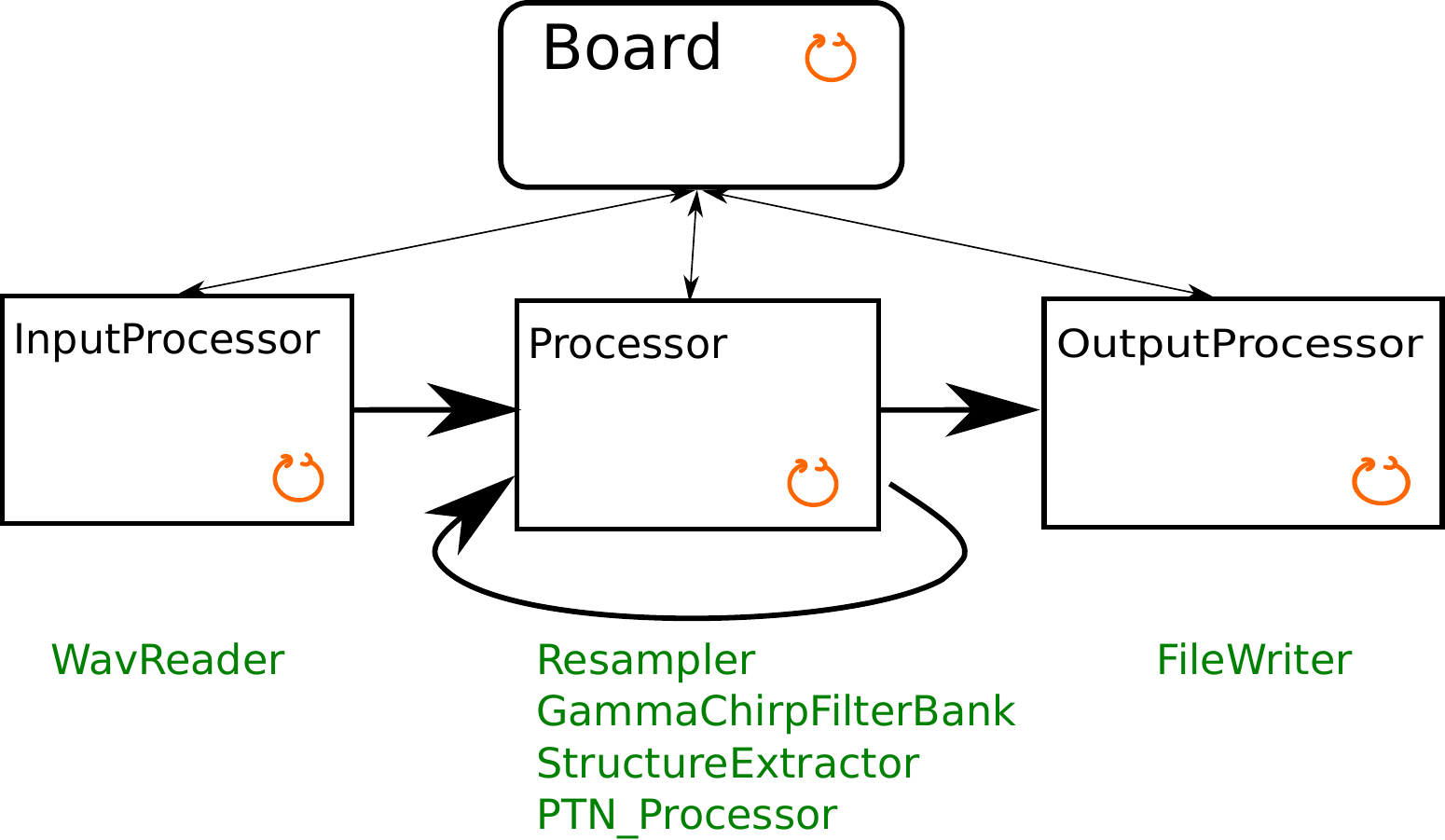}
\caption{File Processing: A sound signal is produced by an
InputProcessor (in our file processing example WavReader), then enters a
network of processing steps encoded in Processors. In order in which
they produce their output Resampler, GammaChirpFilterbank,
StructureExtractor, PTN\_Processor. The network can contain forks and
merges but no loops because of our alignment method. Output from the
Processors can be written to output by OutputProcessors (in our file
processing example FileWriter). In our example the cochleogram computed
in the GammaChirpFilterbank is passed to both the StructureExtractor and
the PTN\_Processor. And a merge occurs because the StructureExtractor
output is also passed the PTN\_Processor where it is merged with the
cochleogram. The use of pipes and the direction of data flow is
indicated by arrows.\label{fig:file_diagram}}
\end{figure}

Figure \ref{fig:file_diagram} shows how the processors are embedded in
the streamboard architecture. The board is the process from which the
different processors are spawned. Processors running on the same machine
are linked through pipes.

\newpage

\begin{figure}[htbp]
\centering
\includegraphics[width=.7\textwidth]{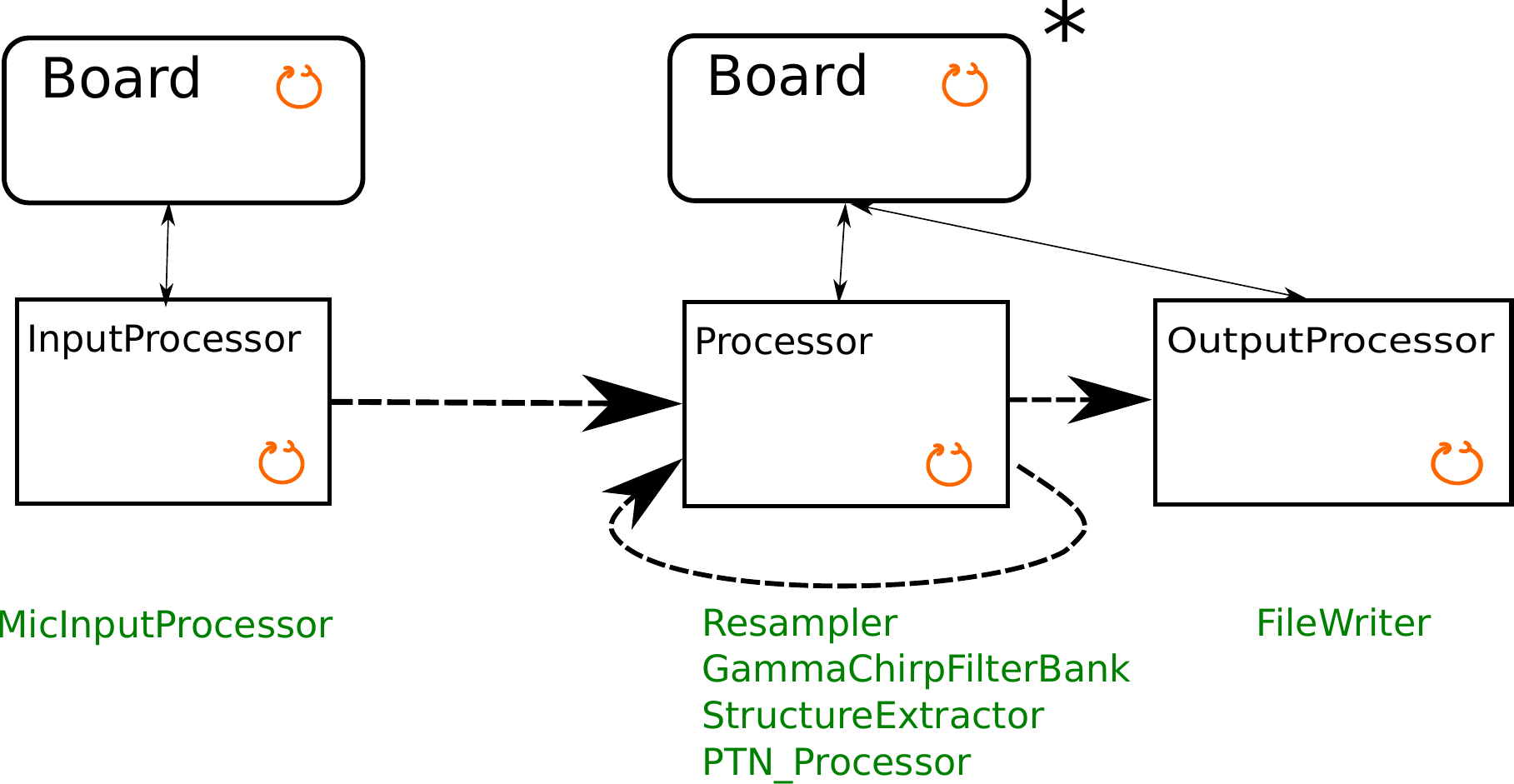}
\caption{Live Microphone Input Processing: A sound signal is produced by
an InputProcessor, in this use case an instance of MicInputProcessor,
then enters a network of processing steps encoded in Processors. In the
order in which they produce their output Resampler,
GammaChirpFilterbank, StructureExtractor, PTN\_Processor. The network
can contain forks and merges but no loops because of our alignment
method. Output from the Processors can be written to output by
OutputProcessors here an instance of FileWriter. In our example the
cochleogram computed in the GammaChirpFilterbank is passed to both the
StructureExtractor and the PTN\_Processor. And a merge occurs because
the StructureExtractor output is also passed to the PTN\_Processor where
it is merged with the cochleogram. The use of network connections is
indicated by the dashed lines with an arrowhead indicating the direction
of the data flow. Pipes are still used for communication between board
and processors as a consequence processors located on different machines
are connected to different boards.\label{fig:mic_diagram}}
\end{figure}

Figure \ref{fig:mic_diagram} shows how the processors are embedded in
the streamboard architecture when each of them is running on a separate
machine. The board is again the process from which the different
processors are spawned. However for each machine a separate board is
required to start the processors on that machine.

\end{document}